\documentstyle[preprint,aps,psfig]{revtex}

\def\Journal#1#2#3#4{{#1} {\bf #2}, #3 (#4)}

\def\PRL{\em Phys. Rev. Lett.}

\newcommand{\beq}{\begin{equation}}
\newcommand{\eeq}{\end{equation}}
\newcommand{\beqs}{\begin{eqnarray}}
\newcommand{\eeqs}{\end{eqnarray}}

\def\lsim{\ \rlap{\raise 3pt \hbox{$<$}}{\lower 3pt \hbox{$\sim$}}\ }
\def\gsim{\ \rlap{\raise 3pt \hbox{$>$}}{\lower 3pt \hbox{$\sim$}}\ }

\begin{document}
\textheight = 23.2cm

\draft
{\tighten 
\preprint{\vbox{\hbox{IFIC/01-15}
                \hbox{hep-ph/0104054}
                \hbox{April 2001}
}}
  \title{~ \\ Testing the principle of equivalence by \\
  supernova neutrinos}
\author{ 
M.\ M.\ Guzzo$^{1}$, 
H.\ Nunokawa$^{1}$ and 
R. Tom\`as$^{2}$
}
\address{\sl ~ \\
$^1$ Instituto de F\' {\i}sica Gleb Wataghin \\
     Universidade Estadual de Campinas, UNICAMP \\    
     13083-970 Campinas SP, Brazil \\ 
\vspace{3mm}
$^2$ Institut de F\'\i sica Corpuscular (IFIC), CSIC - U.
   de Val\`encia, \\  Edifici Instituts d'Investigaci\'o, 
   Apartado de Correos 22085\\ 
     E-46071--Val\`encia, Spain}
\maketitle
\vspace{.5cm}
            
\hfuzz=25pt

\begin{abstract}

\noindent
We study the possible impact of the neutrino oscillation 
which could be induced by a tiny violation of equivalence 
principle (VEP) for neutrinos emitted from supernova
driven by gravitational collapse. 
Due to the absence of any significant indication of 
neutrino oscillation in the SN1987A data, we obtain 
sever bounds on relevant VEP parameters 
$\delta \gamma\lsim O(10^{-31}$) for massless or 
degenerated neutrinos and 
$\delta \gamma \lsim O(10^{-16})\times [\Delta m^2/10^{-5}$ eV$^2$]
for massive neutrinos. 

\end{abstract}

} 

\section{Introduction}

Principle of equivalence, one of the fundamental 
bases in Einstein's general relativity, 
has been tested for many years with great accuracy.  
One can state the possible violation of equivalence 
principle (VEP) in several ways. 
For massive objects, existence of VEP implies that 
gravitational mass is not equal to inertial mass. 
For macroscopic objects, by the torsion balance experiment, 
E\"otv\"os and his collaborators \cite{EPF} obtained 
the stringent bound $\eta \lsim 3 \times 10^{-9}$, 
where $\eta$ is the parameter which measures 
the difference between gravitational and inertial mass. 
Forty years later, Dicke {\it et al.}\cite{RKD} 
improved the bound to $\eta \lsim 3 \times 10^{-11}$
by removing some systematic uncertainties 
making use of the influence of Sun's gravitational 
field which could produce a torque with 24 hour periodicity
in the experiment in the presence of VEP. 
With similar apparatus, Braginski and Panov\cite{BP} 
obtained the bound $\eta \lsim 0.9 \times 10^{-12}$. 

Equivalence principle has been tested also for microscopic 
objects. 
Using the neutron free fall refractometry, 
the bound $\eta < 3 \times 10^{-4}$ is obtained\cite{neutron}
for inertial and gravitational mass difference of neutrons. 

For particles like photon and neutrinos, one can not perform 
such a comparison of inertial and gravitational masses.  
One can use, instead, 
another parameter, $\delta \gamma$, in order to test VEP, 
which measures the difference of the gravitational 
coupling between the two particles involved, 
$\delta \gamma_{\beta\alpha} \equiv (G_\beta-G_\alpha)/(G_\beta+G_\alpha)$. 
A mild bound $|\delta \gamma_{\gamma \nu}| \lsim 10^{-3}$ 
for violation of the equivalence principle for photons and 
neutrinos have been derived \cite{LKT} by using the small 
arrival time differences between photons and neutrinos 
from the supernova SN1987A observed by Kamiokande
and IMB detectors~\cite{SN1987A}. 

For neutrinos, more sensitive test of VEP can be performed by 
using neutrino oscillation which can be induced by VEP even if 
neutrinos are massless~\cite{gasperini,hl}. 
Based on this proposal, many works have been performed
~\cite{Butler,pantaleone,form,kuo,gnz,Casini,vep_MN,MS,fly,FLMS,IMY,Mann,pkm,Horvat,BCDM}. 
In Ref.\ \cite{glashowetal} such VEP induced neutrino 
oscillation was shown 
to be phenomenologically equivalent to 
neutrino oscillations induced by a possible 
violation of Lorentz invariance~\cite{cg1}. 

Some theoretical insight on the type of gravitational 
potential that could violate the weak equivalence principle 
can be found in Ref.\ \cite{amn}. 
Some discussions on the possibility that 
violation of equivalence principle comes from 
string theory are found in Ref.\ \cite{HL98}. 
A discussion on the departure from exact Lorentz invariance in 
the standard model Lagrangian  is developed in Ref.\ \cite{cg2}.

A possibility to test VEP by the future long-base line neutrino 
oscillation experiment has been discussed in Ref.~\cite{IMY}
whereas some bounds from the current accelerator based 
neutrino oscillation experiments were obtained in Ref.~\cite{pkm}. 
Recently such neutrino oscillation mechanisms have been 
investigated ~\cite{fly,FLMS}  in the light of the experimental 
results from Super-Kamiokande (SK) on the atmospheric 
neutrino anomaly, obtaining stringent limits 
for the $\nu_\mu \to \nu_\tau$ channel.

The possibility of solving the solar neutrino problem (SNP) 
by these gravitationally induced neutrino oscillations 
have been investigated in Refs.~\cite{hl,Butler,pantaleone,form,kuo,gnz,Casini}. 
On the other hand, possibilities to constrain VEP parameters 
by using astrophysical neutrino sources are discussed in 
Ref.~\cite{vep_MN} for solar neutrinos and in 
Ref.~\cite{MS} for very high energy neutrinos from AGN. 
The possible effect of neutrino oscillation induced by 
the violation of equivalence principle (VEP) as an explanation 
of the pulsar kick velocity has been considered in Refs.~\cite{Horvat,BCDM}. 
In Ref.~\cite{form}, it was discussed that supernova (SN) explosion 
dynamics  as well as heavy element nucleosynthesis can be 
significantly affected by VEP induced neutrino oscillation. 

In this paper, extending the discussions in Ref.~\cite{form}, we 
consider the possible impact of VEP induced neutrino oscillation 
for SN neutrinos, in particular for the observation 
of $\bar{\nu}_e$ signal. 
In this work, we do not consider the possible effect of 
neutrino oscillation induced by VEP 
on heavy element nucleosynthesis in SN~\cite{qian}  
as the bounds on VEP parameters from this effect 
is expected to be weaker. 
We show that one can try to constrain the relevant VEP 
parameters using the observed anti-electron neutrinos 
events from supernova SN1987A following the same argument 
used in Refs. \cite{SSB94,JNR96}. 
In Sec. II, we discuss the formalism we use in this work. 
In Sec. III we describe some features of SN neutrinos and 
discuss how to use supernova neutrino data
to constrain oscillation parameters. 
In Sec. IV we present our results. 
In Sec. V we give conclusions. 
\section{Neutrino propagation in the presence 
of violation of equivalence principle}

Here we describe the framework we will use in this work. 
Let us assume that neutrinos of 
different species are not universally coupled to gravity, 
or neutrinos of different flavor propagate 
with different gravitational red shift in a given 
gravitational field. 
In such case, the weak interactions eigenstates $|\nu_W\rangle$ 
and gravitational eigenstates $|\nu_G \rangle$, the bases with which 
gravitational part of the energy is 
diagonalized, can, in general, be different, 
$|\nu_W\rangle \ne |\nu_G \rangle$. 
If neutrinos are massive, 
the mass eigenstates  
$|\nu_M \rangle$, can be different from both of these two states, 
$|\nu_M \rangle \ne |\nu_W\rangle \ne |\nu_G \rangle$. 
For two neutrino system, $\nu_e-\nu_x$ ($x=\mu$ or $\tau$), 
we can describe such a general situation as follows, 
\begin{eqnarray}
\left[\matrix{\nu_e \cr\    \nu_{\mu} \cr}\right] 
&=&
\left[\matrix{
\cos\theta_G & \sin\theta_G \cr
-\sin\theta_G & \cos\theta_G \cr}
\right]
\left[\matrix{\nu_{G1} \cr\    \nu_{G2} \cr}\right]\nonumber,\\
&=&
\left[\matrix{
\cos\theta_M & \sin\theta_M \cr
-\sin\theta_M & \cos\theta_M \cr}
\right]
\left[\matrix{\nu_{M1} \cr\    \nu_{M2} \cr}\right],
\label{eqn:matrix}
\end{eqnarray}
where $\theta_G$ is the mixing angle due to the presence of 
VEP which is in general different from the usual flavor mixing 
$\theta_M$ between weak and mass eigenstates, $\theta_G \ne \theta_M$,
implying $|\nu_M \rangle \ne |\nu_G \rangle$. 

For simplicity, let us consider, the system of two neutrino 
flavors. 
The evolution equation of massive neutrino
for one generation without flavor mixing 
in the gravitational field in the supernova 
can be described by, 
\begin{equation} 
i\frac{\text{d}}{\text{d}r} \nu 
= E\left[ 1 - \frac{m^2}{2E^2}-\phi(r) \right]\nu,
\label{evol}
\end{equation} 
where the gravitational potential $\phi(r)$ 
describes the red shift of neutrinos 
and it is given, up to the first order in 
Newton's gravitational constant $G$, 
assuming the spherically symmetric metric and 
taking the center of the star as origin of 
the coordinate, as follows, 

\begin{equation} 
\phi(r) = -G\left[ \frac{{\cal M}(r)}{r} 
+ \int_r^\infty \frac{{\cal M}(r)}{r^2}dr
\right] +\phi_{sc}, 
\end{equation} 
where 
\begin{equation} 
{\cal M}(r) \equiv 
\int_0^r \text{d}r' 4 \pi r'^2 \rho(r')
\end{equation} 
is the total mass contained 
in the volume with radius $r$, and 
$\phi_{sc}$ is the background potential which 
can be taken as the one coming from the local 
Super-cluster~\cite{kenyon}. 
For definiteness, in this work, we take 
$\phi_{sc} = - 3 \times 10^{-5}$, the value estimated 
in Ref.~\cite{kenyon}.

It is straight forward to generalize the evolution 
equation (\ref{evol}) for more than one generation with 
mixing. 
Since, as we will discuss in detail in the following sections, 
we are most interested in the oscillation effect for 
anti-electron neutrinos, $\bar{\nu}_e$, from now on we 
will consider only 
$\bar{\nu}_e-\bar{\nu}_x$ ($x=\mu$ or $\tau$) 
system unless otherwise stated. 
The evolution equation for massive anti-neutrinos in the presence of 
VEP can be written as follows \cite{gasperini},
\begin{equation} 
\hskip  -0.4cm
i{\displaystyle{d \over \displaystyle{dt}}
\left[ \begin{array}{c} \bar{\nu}_e \\
\bar{\nu}_x \end{array} \right] 
} 
= 
\left[ \begin{array}{cc} 
-V_{{e}} - V_G\cos 2\theta_G -\Delta\cos 2\theta_M
&  \frac{1}{2}(V_G\sin 2 \theta_G 
+ \Delta\sin 2 \theta_M) \\
\frac{1}{2}(V_G\sin 2 \theta_G 
+ \Delta\sin 2 \theta_M)
& 0 \end{array} \right] 
\left[ \begin{array}{c} \bar{\nu}_e \\
\bar{\nu}_x \end{array} \right], 
\label{oscila} 
\end{equation} 
where $V_{e} \equiv \sqrt{2}G_F \rho Y_e/m_N$ is the
standard matter potential, 
$\Delta \equiv \Delta m^2/2E $, 
$\theta_M$ is the usual flavor mixing induced by mass
which can be different from $\theta_G$ 
and 
$V_G(r)  \equiv 2\delta \gamma E \phi(r)$.  
For neutrinos system, ${\nu}_e-{\nu}_x$, 
the same equation hold by replacing $-V_e$ by $V_e$. 

We assume that as a reasonably good approximation 
the density profile in the relevant region of the 
star takes the following form, 
\begin{equation} 
\rho(r) = \rho_0 \left(\frac{R_0}{r}\right)^m,
\end{equation} 
where $m=5-7$ in the inner part and $m=3$ in the outer 
layer of the star. 
For definiteness, we take 
$\rho_0 =\rho_1 = 8\times 10^{14}$ g/cm$^3$ and 
$R_0 = R_1 = 10^6$ cm for the inner part and 
$\rho_0 = \rho_2 = 4\times 10^4$ g/cm$^3$ and 
$R_0 = R_2 = 10^9$ for the outer part of the star. 
The radius which defines inner/outer part has been taken to be the 
distance at which both definitions merge, $R_{12}=4.5\times 10^6$ cm.

Under above assumptions, $\phi(r)$ can be explicitly computed as, 
\begin{equation} 
\phi(r) = -G\frac{M_1}{r}
\left[5-2 \left(\frac{R_1}{r}\right)^2\right]
-2 \frac{GM_2}{R_2} + \phi_{sc} \ \ \ 
\text{for} \ R_1 < r < R_{12}, 
\end{equation} 

\begin{equation} 
\phi(r) = -G\frac{M_1}{r}
\left[5 + 6 \left(\frac{R_1}{R_2}\right)^2
\ln\left(\frac{r}{R_2}\right)
\right]  + \phi_{sc} \ \ \ 
\text{for} \ 
R_{12}  < r, 
\end{equation} 
where $M_1 \equiv 4 \pi\rho_1 R^3_1/3$ and 
$M_2 \equiv M_1(R_1/R_2)^2$. 
In Fig. 1 we show the profiles of 
$V_e + \Delta \cos2 \theta_M$ and 
$V_G$ in eV for some values of mass and mixing parameters, 
and VEP parameters, 
as a function of distance from the center of
the star.  

A resonantly enhanced conversion can occur if the 
the resonance condition as well as the adiabaticity 
conditions are satisfied. 
The resonance condition is given by, 
\begin{equation} 
V_e = -  V_G\cos 2\theta_G -  \Delta\cos 2\theta_M,
\end{equation} 
Note that depending on the sign of $V_G$ and $\Delta$, 
resonance condition can be satisfied either for 
neutrino or anti-neutrino.

The adiabaticity parameter $\kappa(r)$ at position $r$ 
can be defined as
\begin{equation} 
\kappa(r) 
\equiv {\Delta H(r)} \left|\frac{\text{d}\theta_m}{\text{d}r}\right|^{-1}, 
\label{adiabaticity} 
\end{equation} 
where $\Delta H(r)$ is the energy splitting between two energy 
eigenvalues of the Hamiltonian, 
$\Delta H(r) \equiv E_{m2} - E_{m1}$. 
Explicitly, $\kappa(r)$ can be written as follows,  
\begin{equation} 
\kappa(r) 
= 
\frac{\left(\sin2\theta_M+\frac{V_G}{\Delta}\sin2\theta_G\right)^2}
{|\sin^32\theta_m|} 
\frac{4\pi \ell_0 \ell^{eff}_\rho}{\ell^2_\nu},
\end{equation} 
where
$\ell_0 \equiv 2\pi m_N/ \sqrt{2}G_F \rho Y_e$ is the
matter refraction length, 
$\ell_\nu \equiv 4\pi E/\Delta m^2$ is the 
oscillation length in vacuum without VEP and  
$\ell^{eff}_\rho$ is defined as, 
\begin{equation} 
\ell^{eff}_\rho \equiv 
\left| \ell^{-1}_G
\frac{
\sin 2\theta_G+\frac{\Delta}{V_e}\sin2(\theta_G-\theta_M)}
{\sin 2\theta_G+\frac{\Delta}{V_G}\sin2\theta_M } 
-\ell^{-1}_\rho
\right|^{-1},
\end{equation}
where 
$\ell_\rho \equiv |\text{d} \ln V_e/\text{d}r|^{-1}$ and 
$\ell_G \equiv |\text{d} \ln V_G/\text{d}r|^{-1}$ 
are the density and gravitational scale height, 
respectively. 
Mixing angle in matter $\theta_m$ in the presence of VEP 
is given by,
\begin{equation} 
\tan 2\theta_m = 
\frac{2H_{12}}{H_{22}-H_{11}}
= \frac{\Delta \sin{2\theta_M} + V_G \sin{2\theta_G}}
{V_e + \Delta \cos{2\theta_M} + V_G \cos{2\theta_G}}. 
\end{equation} 

For the case where resonant conversion occur, at resonance point, 
$\kappa$ takes somewhat simpler form, 
\begin{equation} 
\kappa(r_{res}) 
= 
2\frac{(\Delta \sin2\theta_M + V_G(r_{res}) 
\sin 2\theta_G)^2}
{\left| \frac{\text{d}V_e}{\text{d}r} 
+ \frac{\text{d}V_G}{\text{d}r} \cos 2\theta_G 
\right|_{res}}
\end{equation} 
from which it is possible to derive the known cases, 
when $\Delta m^2 \rightarrow 0$ or $\delta \gamma \rightarrow 0$.

We compute the survival probability of $\bar{\nu}_e$ 
at the surface of the outer region of the supernova 
$P^{SN}_{ee}$, 
by using the following formula, 
\begin{equation}
P^{SN}_{ee} \equiv 
P^{SN}(\bar{\nu}_e \to \bar{\nu}_e) = 
\frac{1}{2} + 
(\frac{1}{2} - P_{c})
\cos2\theta_i \cos2\theta_f,
\end{equation}
where $ \theta_i$ and  $\theta_f$ are
the initial and final mixing angle
which 
can be taken as $\cos2\theta_i = 1$ 
and $\cos2\theta_f = \cos2\theta_0$ where $\theta_0$ is 
defined as
\begin{equation} 
\tan 2\theta_0 \equiv 
\frac{\Delta \sin{2\theta_M} + V_G^{vac} \sin{2\theta_G}}
{ \Delta \cos{2\theta_M} + V_G^{vac}  \cos{2\theta_G}}. 
\label{newangle}
\end{equation} 

The crossing probability $P_c$ is computed as~\cite{petcov88}, 
\begin{equation} 
P_{c} = 
\frac{\exp(-\frac{\pi}{4} \kappa F)-
\exp(-\frac{\pi}{4} \kappa F/\sin^2\theta_0)}
{1-\exp(-\frac{\pi}{4} \kappa F/\sin^2\theta_0)},
\end{equation} 
where, as an approximation, following Ref.~\cite{petcov88},  
we use $F \equiv | 1-\tan^2\theta_0| $ 
which is valid for exponentially decreasing 
density profile. 
When $\bar{\nu}_e$ does not undergo resonance, 
as an approximation, we use the same formulas but 
with $\kappa$ estimated at the resonance point 
for ${\nu}_e$ channel. 

We compute the final survival probability at the detector as 
follows\cite{DLS}, 
\begin{equation}
P_D(\bar{\nu}_e) 
= \frac{P^{SN}_{ee} - \sin^2 \theta_{0} + P^E_{{2e}} (1-2 P^{SN}_{ee})}
{\cos2\theta_{0}} + P_{coh},
\label{p_detector}
\end{equation}
where $P^{SN}_{ee}$ is the averaged 
survival probability of $\bar{\nu}_e$ at the 
surface of supernova, 
computed by taking into account the energy 
spectrum (see eq. (\ref{ene_spec}))
and 
$P^E_{2e}$ is the conversion probability of 
$\bar{\nu}_2 \to \bar{\nu}_e$ after crossing the Earth.   
We define new eigenstates, $\nu_1$ and $\nu_2$,
which are neither the mass nor gravitational eigenstates 
in the presence of mass and VEP, by using the
mixing angle defined in eq. (\ref{newangle}) as follows, 
\begin{equation}
\left[\matrix{\nu_e \cr\    \nu_{\mu} \cr}\right] 
= 
\left[\matrix{
\cos\theta_0 & \sin\theta_0 \cr
-\sin\theta_0 & \cos\theta_0 \cr}
\right]
\left[\matrix{\nu_{1} \cr\    \nu_{2} \cr}\right]
\label{eqn:neweigenstates}
\end{equation}

$P_{coh}$ is the ``coherent'' part of the probability 
which is given by, 
\begin{equation}
 P_{coh} = 
2  \sqrt{P^{SN}_{e1}(1-P^{SN}_{e1})P^{E}_{e1}(1-P^{E}_{e1})
} \cos(\alpha_{SN}+\alpha_E+\alpha),
\end{equation}
where 
$P^{SN}_{e1}$ and $P^{E}_{e1}$ denote the conversion probability 
of $\bar{\nu}_e \to \bar{\nu}_1$ in the supernova
and in the Earth, respectively, 
$\alpha_{SN}$ 
is the relative phase difference between 
the $\bar{\nu}_1$ and $\bar{\nu}_2$ states at the surface
of the supernova, 
$\alpha_E$
is the phase difference 
between the amplitude of $\bar{\nu}_1\to\bar{\nu}_e$ 
and $\bar{\nu}_2\to\bar{\nu}_e$ after crossing the Earth, 
and 
\begin{equation}
\alpha
= \left[\Delta^2 + (\delta \gamma \phi_{sc})^2
+ \frac{1}{2} 
\Delta\cdot \delta \gamma \phi_{sc} \sin{4\theta_G} \sin{4\theta_M}
\right]^{\frac{1}{2}} L, 
\label{phi}
\end{equation}
where $L$ is the distance between supernova and the Earth. 
In our analysis we will use $L=50$ kpc for the distance 
between SN1987A and the Earth. 
We note that $ P_{coh}$ could be important if sum of 
these phases is not so large. 
If one of them is very large, then the coherent term will 
be averaged out, after we take into account the energy spread
of supernova neutrinos. 
The largest contribution is coming from $\alpha$ and 
when $\alpha$ is small, the other phases are expected 
to be very small compared to $\alpha$, and therefore, 
we can simply neglect other two phases 
$\alpha_{SN}$ and $\alpha_{E}$, 
as a good approximation, to compute $ P_{coh}$. 

Assuming that the density is constant, 
$P^E_{2e}$ can be computed as follows~\cite{SSB94}, 
 
\begin{equation}
P^E_{2e} \equiv P^E(\bar{\nu}_2 \to \bar{\nu}_e)= \sin^2 \theta_0 
+ \sin 2\theta_\oplus \sin 2(\theta_\oplus -\theta_0) 
\sin^2\left(\pi\frac{\ell}{\ell_\oplus}\right),
\end{equation}
where $\ell$ is the distance traveled by neutrino
inside the Earth, 
$\theta_\oplus$ and $\ell_\oplus$ are the mixing angle 
and the neutrino oscillation length, respectively, 
in the Earth. 

\section{Possible impact of oscillation for supernova neutrinos}

Let us consider the influence of neutrino oscillation 
between $\bar{\nu}_e$ and $\bar{\nu}_x$ ($x = \mu$ or $\tau$)
for SN neutrinos. 
From the view point of observation, influence on $\bar{\nu}_e$ 
signal is most important because the cross section for 
absorption reaction $\bar{\nu}_e p \to e^+ n$ is much larger than 
elastic scattering processes $\nu_x e^- \to \nu_x e^-$ ($x=e,\mu,\tau$). 
In fact it is considered that observed events at Kamiokande-II 
and IMB detectors~\cite{SN1987A} from 
SN1987A are induced by $\bar{\nu}_e$. 

The energy spectrum of supernova neutrinos can be 
described by ``pinched" Fermi-Dirac distribution 
where pinched form can be parametrized by 
the ``effective'' chemical potential $\eta_i$ as follows
~\cite{JH89}, 
\begin{equation}
f_i(E_\nu) \propto
\frac{E^2_\nu}{\exp(E_\nu/T_i-\eta_i)+1},
\label{ene_spec}
\end{equation}
where $T_i$ and $\eta_i$ are the temperature 
and effective chemical potential, respectively, 
for $i$-th neutrino species. 
Because of the different interaction rates of 
neutrinos in the proto-neutron star, it is expected that there 
exist the following hierarchy relation between temperatures, 
\begin{equation}
T_{\nu_e} < T_{\bar{\nu}_e} < T_{\nu_{\mu,\tau}} 
\simeq  T_{\bar{\nu}_{\mu,\tau}}.
\end{equation}
Due to such temperature difference, 
average neutrino energies in typical numerical SN simulations, 
which are approximately related with temperatures 
as $\langle E_{\nu}\rangle \approx 3 T_\nu$, 
are~\cite{MWS,SNsimu,Janka}, 
\begin{eqnarray}
\langle E_{\nu_e}\rangle & \approx& 11-12\ \mbox{MeV},\nonumber\\
\langle E_{\bar\nu_e}\rangle &\approx& 14-17\ \mbox{MeV},\nonumber\\
\langle E_{\nu_{\tau(\mu)}}\rangle & \approx&  \langle
E_{\bar\nu_{\tau(\mu)}}\rangle\approx 24-27\ \mbox{MeV}.
\label{ene_ave}
\end{eqnarray}
Such energy hierarchy is of crucial importance for our 
discussion as we will see below. 

If neutrino oscillation between $\bar{\nu}_e$ and 
$\bar{\nu}_x$ ($x=\mu,\tau$) occurs in SN
and/or in the space between SN and Earth, 
expected neutrino flux at the Earth can be given by, 
\begin{equation}
F_{\bar{\nu}_e}
= (1-p) F^0_{\bar{\nu}_e} + p F^0_{\bar{\nu}_x},
\label{phi_obs}
\end{equation}
where $p$ is the permutation factor, or the energy averaged 
conversion probability for $\bar{\nu}_e \leftrightarrow \bar{\nu}_x$  
which can be computed by convoluting 
the probability in eq. (\ref{p_detector}) 
in the previous section with the Fermi-Dirac 
energy distribution in eq. (\ref{ene_spec}), 
and $F^0_{\bar{\nu}}$ denotes the time integrated original 
neutrino flux in the absence of oscillation. 
For simplicity, we will set the ``effective chemical 
potential, $\eta_i$, in eq. (\ref{ene_spec}), equal to zero
in this work. 

Because of the energy hierarchy in eq. (\ref{ene_ave}), 
large oscillation between 
$\bar{\nu}_e \leftrightarrow \bar{\nu}_{\mu,\tau}$ 
would make the observable energy of $\bar{\nu}_e$ larger
than the theoretical predictions. 
However, the observed SN1987A neutrino data by 
Kamiokande and IMB detectors~\cite{SN1987A} imply rather 
smaller average energy, which are in the lower side of 
the theoretical expectations, indicating that there was 
no significant oscillation between $\bar{\nu}_e$ and 
$\bar{\nu}_{\mu,\tau}$.  
Based on this argument, it was obtained in Ref.~\cite{SSB94}
the bound on permutation parameter as $ p < 0.35$ 
at 99 \% C. L., which disfavors large oscillation between
$\bar{\nu}_e$ and  $\bar{\nu}_{\mu,\tau}$. 
The same argument was used in Ref.~\cite{JNR96} 
and most recently in Refs.~\cite{LS00,KTV,MN00,MY00,DC00}, 
to constrain large oscillation for neutrinos from SN1987A. 

%
\section{Bounds on VEP parameters from supernova}

In this section we discuss our results which are obtained 
using the formulas described in the previous section. 

\subsection{massless neutrinos case}
First let us discuss the case without neutrino mass, or the
case where masses are degenerated. 
In this case, the relevant formulas can be obtained 
by putting $\Delta m^2 \to 0$ in the previous section. 
In order to cover all the physical case, we fix 
$\delta \gamma$ to be positive but consider 
$\theta_G$ in the range from 0 to $\pi/2$, 
analogous to the case of usual MSW effect~\cite{darkside}. 
Here we try to constrain the VEP parameters
from the observation of the supernova SN1987A 
neutrinos by Kamiokande and IMB following the
argument discussed in Refs. \cite{SSB94,JNR96}.

In Fig.\ \ref{fig2} we show the 
iso-contours of permutation factor for anti electron neutrinos
from supernova SN1987A on the $\tan^2 \theta_G-\delta \gamma$ 
plane. 
For parameters in the inner region of these 
curves, the neutrino conversion is efficient so that 
the energy spectrum of $\bar{\nu}_e$ gets harder which 
would lead to worse agreement of the data with 
the predictions from SN simulations. 
If we use the criteria adopted in  Ref. \cite{SSB94}, 
shaded region could be disfavored at 99 \% C.L. 
In the same plot, for the purpose of comparison, we also show 
the region (inside the dashed curves) which can be excluded 
by the laboratory experiments obtained in Ref.~\cite{pkm}. 

We note that for $\delta \gamma$ larger than $10^{-20}$, 
strong VEP induced resonant conversion~\cite{hl},
which is similar to the well known MSW effect, 
can occur inside the supernova, and the 
permutation parameter can be larger than 0.5
even for very small values of gravitational mixing angle 
$\theta_G$. We see that significantly large parameter region 
can be disfavored which can not be tested by laboratory 
experiments (see dashed curve). 

On the other hand, for $\delta \gamma$ smaller than 
$\sim 10^{-20}$, the dominant effect is the 
VEP induced vacuum like-oscillation 
with the probability, 
\begin{equation} 
P(\bar{\nu}_e \to\bar{\nu}_e)
= 1-\sin^2 2\theta_G \sin^2[\delta \gamma \phi_{sc} EL]. 
\end{equation} 
In this case the permutation parameter can be large 
if the mixing angle $\theta_G$ is large 
and $\delta \gamma\lsim 10^{-31}$ and therefore, 
such parameter region could be constrained.  
The reason why we can obtain such stringent bound on 
$\delta \gamma$, 
far below than the existing experimental limit~\cite{pkm}, 
$\delta \gamma \lsim 10^{-19}[|\phi_{sc}|/3\times 10^{-5}]$, 
is simply because of the fact that the distance between 
the observed supernova and Earth 
is much larger (about 50 kpc, for the case of SN1987A)
compared to the baseline of laboratory neutrino 
oscillation experiments or even compared to the baseline 
for solar neutrinos, Sun-Earth distance.

\subsection{massive neutrinos case}

Next let us discuss the case where neutrinos are massive. 
Let us consider how the result presented in the previous 
section can be modified due to mass and usual flavor mixing.  
For the sake of demonstration, we take some particular
sets of mixing parameter implied by the MSW solutions
to the solar neutrino problem~\cite{GHPV}. 
Basically, the main effect is that the sensitivity to VEP 
is weakened due to the presence of usual mass and mixing terms. 
It is easy to see this. When the magnitudes of mass and usual
flavor mixing terms are lager than those by VEP, oscillation 
effect is essentially induced or dominated by the usual mass 
and flavor mixing, and vice versa, as they mainly govern 
the evolution of neutrinos. 

In Fig. 3 we show our results obtained by assuming, 
particular set of the mass and flavor mixing 
implied by the large mixing angle (LMA) 
MSW solution to the solar neutrino problem, 
$\Delta m^2 = 3.2\times10^{-5}$ eV$^2$ 
and $\tan^2  \theta_M = 0.33$ in addition the VEP effect. 
For $\delta \gamma$ smaller than $\sim O(10^{-16})$ the effect of 
neutrino conversion is essentially dominated by the usual 
mass induced oscillation. 
We note that new disfavored region appears for 
$\delta \gamma \sim 10^{-16}-10^{-14}$ for any small 
values of $\theta_G$. This is due to the combined effects of 
the usual mass and VEP induced terms in the neutrino 
evolution Hamiltonian. 
In these parameter regions of $\delta \gamma$ around $\sim O(10^{-16})$,  
resonant conversion is mainly satisfied by the VEP 
parameters whereas the adiabaticity condition is met 
due to the presence of mass induced mixing term or 
$\frac{1}{2}\Delta \sin2\theta_M$, 
even if $\theta_G$ is zero. 

In Fig. 4 we show the similar
plot but for the small mixing angle (SMA) MSW solution
with $\Delta m^2 = 5.0\times10^{-6}$ eV$^2$ 
and $\tan^2 \theta_M = 5.8\times10^{-4}$. 
In this case, unlike in Fig. 3, no new disfavored region appears 
for $\delta \gamma \sim O(10^{-16})$. 
This is because for these values of $\delta \gamma$ with 
small $\theta_G \ll 1$, 
the adiabaticity condition,  which was satisfied in the case of 
LMA solution, is not satisfied anymore 
since $\frac{1}{2}\Delta \sin2\theta_M$ is much smaller 
than the case of LMA. 

Some comments are in order. 
We showed in Fig. 3 and 4 some parameter region which can be
disfavored by SN1987A data assuming the mass and mixing suggested by 
LMA and SMA solutions to the solar neutrino problem. 
We should note, however, that if $\delta \gamma$ is much larger 
than  $\sim 10^{-16}$, solar neutrino flux can be
significantly affected by the VEP effect and hence 
LMA and SMA solutions must be affected accordingly, as 
discussed in Ref.~\cite{vep_MN}. 
Therefore, 
the parameter regions allowed in Fig. 3 and 4 
by the SN1987A data are not necessarily implying 
the parameter region allowed by the solar neutrino data. 

\section{Conclusions}
We have studied the impact of the neutrino oscillation 
which could be induced by a tiny breakdown of the
equivalence principle.
We have obtained regions of parameters which are 
disfavored by the SN1987A neutrino data and showed 
that supernova can prove some region of parameter
space which can not be tested by laboratory experiments. 
We show that supernova neutrinos can be sensitive to 
disfavor very tiny values of $\delta \gamma$ as small
as $O(10^{-31})$, which is far below the 
laboratory bound, if neutrino mass can be neglected.  
We also showed that much smaller gravitationally induced 
mixing angle, 
$\tan^2 \theta_G \ll 10^{-4}$, which can not be tested by 
the laboratory experiments, can be proved by SN neutrinos. 
For massive neutrinos, we showed that the sensitivity 
to these bounds will become worse. 
The effect of VEP in oscillation is lost or weakened
when mass induced terms become larger than VEP terms. 
Let us note that, compared to the bounds on 
VEP obtained by using the macroscopic objects, 
$\eta \lsim 10^{-12}$, mentioned in the introduction, 
these bounds we obtained are much more stringent. 

Let us finally try to compare the sensitivity to VEP parameter 
for neutrinos from supernova with that for neutrinos
from other astrophysical sources. 
With solar neutrinos, it was discussed that 
effect of VEP can be tested for $\delta \gamma$ as small 
as $\sim 10^{-20}$ for massless neutrinos~\cite{gnz} and 
for $\delta \gamma$ as small as $\sim 10^{-16}-10^{-15}$ for 
massive neutrinos if we assume the MSW solutions to the solar 
neutrino problem~\cite{vep_MN}. 
Much better sensitivity can be obtained if we can observe 
very high energy neutrino from AGN~\cite{MS}. 
Due to much higher energy and lager distance, 
it is discussed in Ref.~\cite{MS} that 
one can test VEP for $\delta \gamma$ as small as $10^{-41}$
if neutrinos are massless or degenerated neutrinos
and $\delta \gamma \sim 10^{-28} \times (\Delta
m^2/1 {\rm eV}^2)$ for massive neutrinos. 
However, let us stress that in the absence of such ANG neutrino 
data, currently, SN1987A data seem to provide better bound on 
VEP parameters than any existing terrestrial experiments.

\acknowledgments 

This work was supported by Funda\c{c}\~ao de Amparo \`a
Pesquisa do Estado de S\~ao Paulo (FAPESP) and Conselho Nacional de
Desenvolvimento Cient\'\i fico e Tecnolog\'ogico (CNPq). 
H.N. would like to thank Jose F. W. Valle for his hospitality and 
support during his visit at Univ. of Val\`encia where part of this 
work was done. 
We thank O.L.G.Peres and H. Minakata for useful discussions 
and comments. 
R. T. has been supported by a grant from the Generalitat Valenciana.
R. T. would like to thank the department of cosmic ray at Gleb 
Wataghin physics institute and also P.C. de Holanda and 
M. Queiroz for their hospitality during his visit to UNICAMP.
%

\vglue 0.5cm 
\begin{figure}
\centering\leavevmode
\centerline{\protect\hbox{
\psfig{file=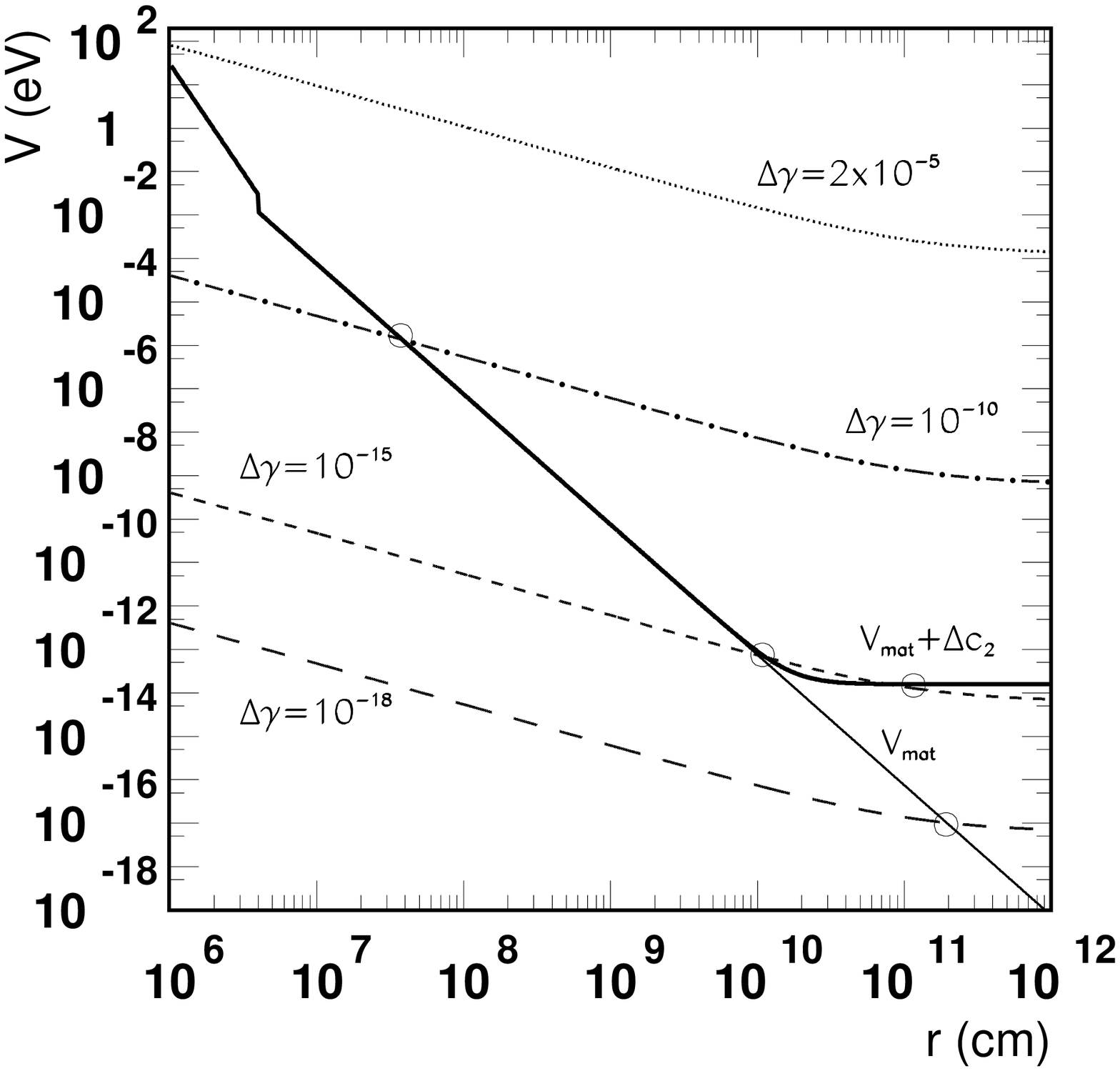,height=13.cm,width=13.cm}
}}
\caption{
$V_G(r)$ are plotted as a function of 
the distance from the center of the star for 
various values of $\delta \gamma$ by dotted, dot-dashed, 
short-dashed, long-dashed curves.  
Also $V_{mat}(r) + \Delta\cos2\theta$ is plotted 
for massless (or degenerate) case 
and $E = 10$ MeV, $\Delta m^2= 10^{-5}$ eV$^2$
and $\sin^2 2\theta_M=0.9$.} 
\label{fig1}
\vglue -10.cm
\end{figure}

\newpage 
\vglue 2cm 
\begin{figure}
\centering\leavevmode
\centerline{\protect\hbox{
\psfig{file=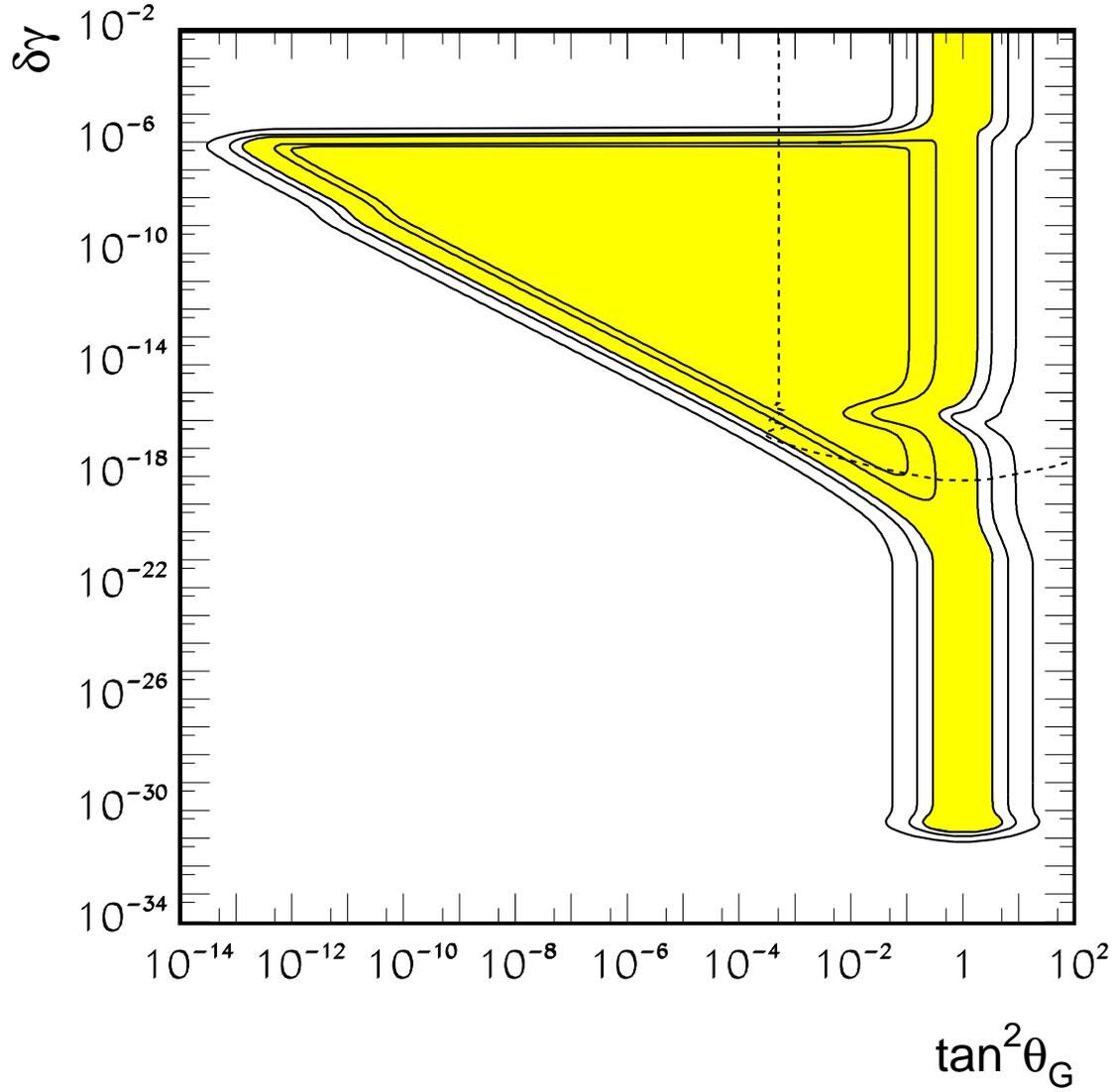,height=16cm,width=16.cm}
}}
\vglue 0.5cm
\caption{
Iso-permutation factor contours for anti electron neutrinos
from supernova on the $\tan^2 \theta_G-\delta \gamma$ plane. 
The curves correspond, from inside to outside, 
$p=0.9,0.75,0.35,0.23,0.1$. 
The shaded region corresponds to $p> 0.35$. 
Inside the dashed curve indicates the excluded region 
obtained from laboratory experiment obtained in 
Ref.~\protect\cite{pkm}. 
}
\label{fig2}
\vglue -0.36cm
\end{figure}

\newpage 
\vglue 2cm 
\begin{figure}
\centering\leavevmode
\centerline{\protect\hbox{
\psfig{file=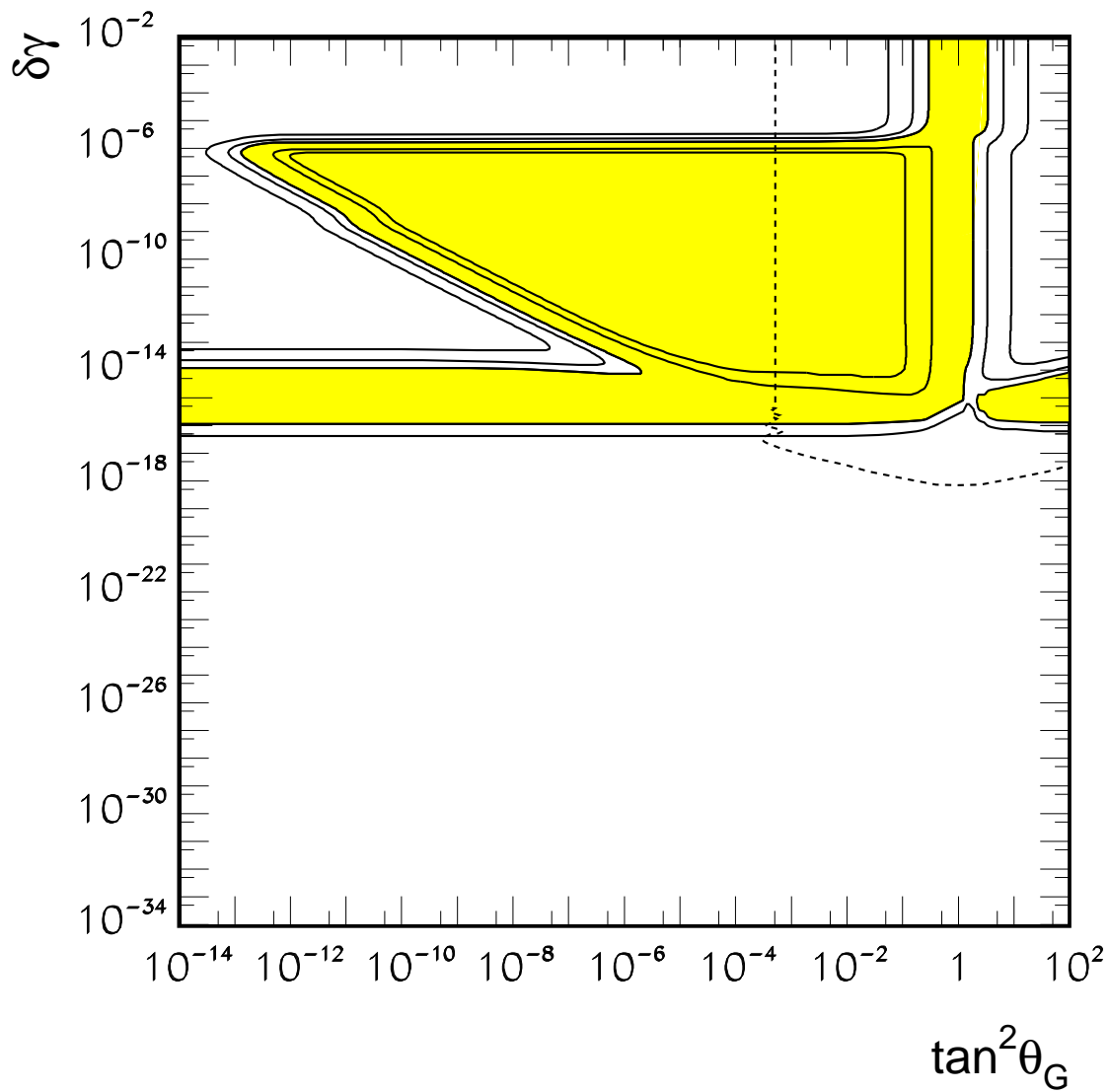,height=16cm,width=16.cm}
}}
\vglue 0.5cm
\caption{
Same as in Fig. 2 but in the presence of mass and 
mixing, $\Delta m^2 = 3.2\times10^{-5}$ eV$^2$ 
and $\tan^2  \theta_M = 0.33$, 
suggested by the LMA MSW solution to the solar
neutrino problem. 
}
\label{fig1}
\vglue -0.36cm
\end{figure}

\newpage 
\vglue 2cm 
\begin{figure}
\centering\leavevmode
\centerline{\protect\hbox{
\psfig{file=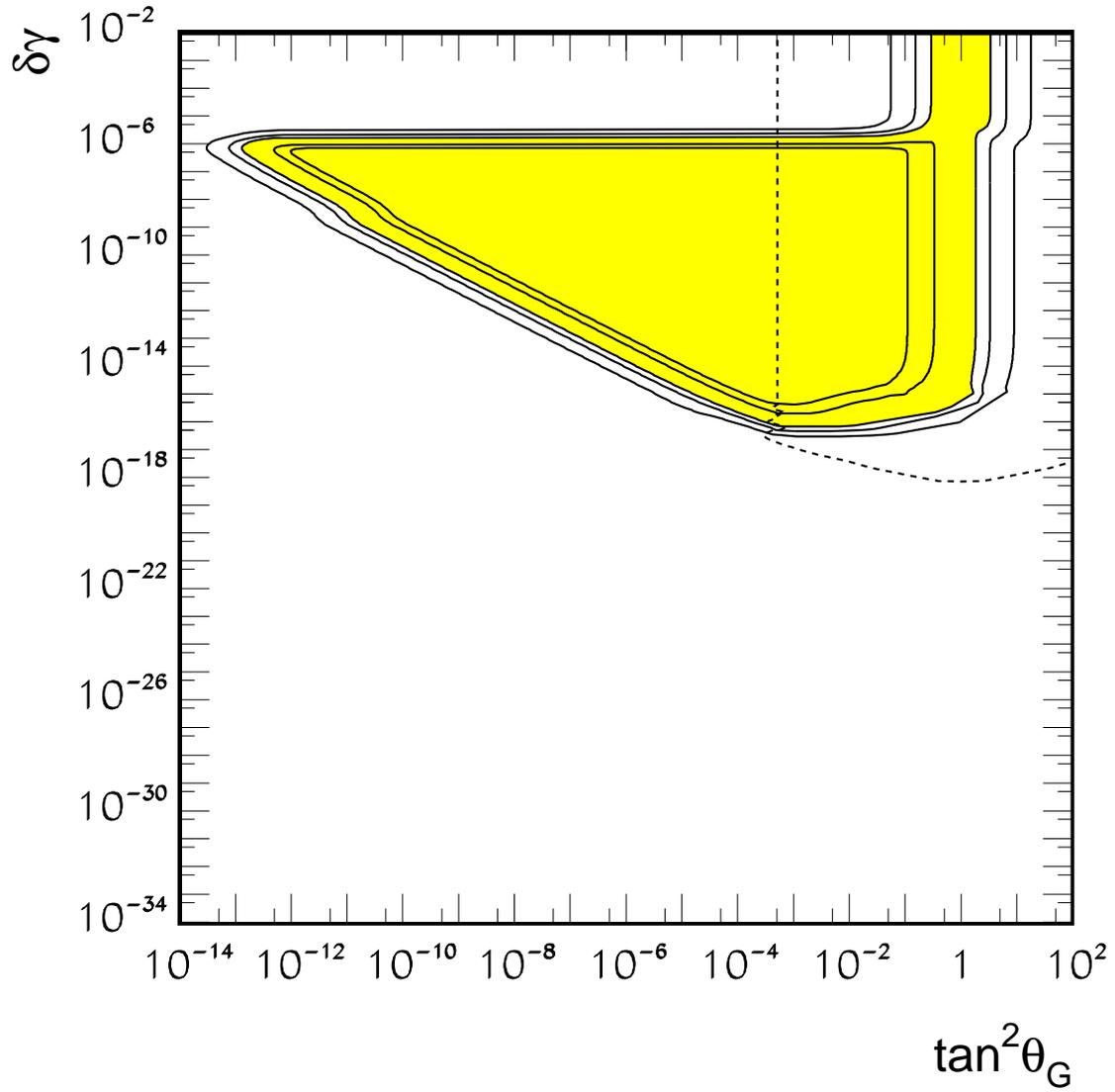,height=16cm,width=16.cm}
}}
\vglue 0.5cm
\caption{
Same as in Fig. 2 but in the presence of mass and 
mixing, $\Delta m^2 = 5.0\times10^{-6}$ eV$^2$ 
and $\tan^2 \theta_M = 5.8\times10^{-4}$, 
suggested by the SMA MSW solution to the solar
neutrino problem. 
}
\label{fig1}
\vglue -0.36cm
\end{figure}

\end{document}